\documentclass[
    ,final            
  ]
  {aipproc}

\layoutstyle{6x9}
\usepackage{graphicx}

\begin{document}

\title{Azimuthal asymmetries from unpolarized data at COMPASS}


\author{C. Schill for the COMPASS collaboration}{
  address={Physikalisches Institut der Universit\"at Freiburg\\Hermann-Herder Str. 3\\ D-79104 Freiburg}
}

\begin{abstract} The investigation of transverse spin and transverse momentum effects in the
nucleon is one of the key physics programs of the COMPASS experiment at CERN. COMPASS investigates
these effects scattering $160$~GeV/c muons off a fixed $NH_3$ or $^6LiD$ target. The azimuthal
asymmetries which appear in the cross-section of semi-inclusive deep-inelastic scattering on an
unpolarized target have been measured. These asymmetries give insight into the intrinsic transverse
momentum of the quarks in the nucleon by the Cahn effect and into a possible correlation between
transverse momentum and transverse spin. New results for azimuthal asymmetries of single hadrons
produced in scattering muons off an unpolarized  $^6LiD$ target are presented. \end{abstract}

\maketitle


\section{Introduction}

The study of transverse momentum dependent distribution functions was initiated by Ralston and
Soper in their study of Drell-Yan processes \cite{Ralston}. In order to understand these
effects in a QCD framework, the description of the partonic structure of the nucleon has been
extended to include the quark transverse spin and transverse momentum $k_T$ \cite{Mulders}.   
The SIDIS cross-section in the one-photon exchange approximation contains
eight transverse-momentum dependent distribution functions. Some of these  can
be extracted in SIDIS measuring the azimuthal distribution of the hadrons in the final state
\cite{Collins}.  The chiral-odd  Boer-Mulders function is of special interest among the 
transverse-momentum dependent distribution functions \cite{Boer}, since it describes the transverse
parton polarization in an unpolarized hadron. The Boer-Mulders function generates azimuthal
asymmetries in unpolarized SIDIS, together with the so-called Cahn effect \cite{Cahn}, which
arises from the fact that the kinematics is non-collinear when $k_T$ is taken into account.

\section{The COMPASS experiment}

COMPASS is a fixed target experiment at the CERN SPS accelerator with a wide physics program
focused on the nucleon spin structure and on hadron spectroscopy \cite{COMPASS}. A $160$~GeV
muon beam is scattered off a transversely polarized $^6$LiD  target. For this analysis data
with opposite target polarization have been spin-averaged, in order to obtain a result for an
unpolarized deuteron target.  The scattered muon and the produced hadrons are detected in a
50~m long wide-acceptance forward spectrometer with excellent particle identification
capabilities \cite{Experiment}. A variety of tracking detectors is used to cope with the
different requirements of position accuracy and rate capability at different angles.

\section{Unpolarized Azimuthal asymmetries in SIDIS}

The cross-section for hadron production in lepton-nucleon SIDIS $\ell N
\rightarrow \ell' h X$ for unpolarized targets and an unpolarized or
longitudinally polarized beam has the following form~\cite{Bacchetta2}:
\begin{equation}
\begin{array}{lcr}\displaystyle
\frac{d\sigma}{dx dy dz d\phi_h dp^2_{h,T}} =
  \frac{\alpha^2}{xyQ^2}
\frac{1+(1-y)^2}{2} \cdot\\[2ex] \displaystyle [ F_{UU,T} + 
  F_{UU,L} + \varepsilon_1 \cos \phi_h F^{\cos \phi_h}_{UU} \\[2ex]
   + \varepsilon_2 \cos(2\phi_h) F^{\cos\; 2\phi_h}_{UU}
   + \lambda_\mu
  \varepsilon_3
  \sin \phi_h F^{\sin \phi_h}_{LU} ]
\end{array}
\end{equation}
where $\alpha$ is the fine structure constant. 
$F_{UU,T}$,  $F_{UU,L}$, $F^{\cos \phi_h}_{UU}$,  $F^{\cos\;
  2\phi_h}_{UU}$ and $F^{\sin \phi_h}_{LU}$ are structure functions. Their 
first and second subscripts indicate the beam and target polarization,
respectively, and the last subscript denotes, if present, the
polarization of the virtual photon.  $\lambda_\mu$ is the 
longitudinal beam polarization and: 
\begin{equation}
\begin{array}{rcl}
\varepsilon_1 & = & \displaystyle\frac{2(2-y)\sqrt{1-y}}{1+(1-y)^2} \\[2ex]
\varepsilon_2 & = & \displaystyle\frac{2(1-y)}{1+(1-y)^2} \\[2ex]
\varepsilon_3 & = & \displaystyle\frac{2 y \sqrt{1-y}}{1+(1-y)^2}
\end{array}
\end{equation}
are depolarization factors.

The Boer-Mulders parton distribution function contributes to  both the $\cos \phi_h$
and the $\cos 2\phi_h$ moments. Another source of  $\cos \phi_h$ and
the $\cos 2\phi_h$ moments in unpolarized scattering is the so-called  Cahn
effect~\cite{Cahn} which arises from the fact that the kinematics is non
collinear when the transverse momentum $k_\perp$  of the quarks is taken into
account. Additionally, perturbative gluon radiation, resulting in higher order
$\alpha_s$ QCD processes, contributes to the observed  $\cos \phi_h$ and the
$\cos 2\phi_h$ moments as well. pQCD effects become important for high
transverse momenta $p_T$ of the produced hadrons.  

In this analysis, data taken with a longitudinally or transversely polarized
$^6$LiD target in the year $2004$ has been spin-averaged in order to obtain
an unpolarized data sample. To select DIS events, kinematic cuts on the negative squared four momentum transfer
$Q^2>1$~(GeV/c)$^2$, the hadronic invariant mass $W>5$~GeV/c$^2$ and the fractional energy
transfer of the muon $0.1<y<0.9$ were applied. A Monte Carlo simulation is used
to correct for acceptance effects of the detector. The SIDIS event generation
is performed by  the LEPTO generator~\cite{lepto}, the experimental setup and the
particle interactions in the  detectors are simulated by the COMPASS Monte Carlo
simulation program COMGEANT. 

The  acceptance of the detector as a function of the azimuthal angle $A(\phi_h)$ is then calculated as the
ratio of  reconstructed over generated events for each bin of $x$, $z$ and $p_T$ in which the
asymmetries are measured. The measured distribution, corrected for acceptance, is fitted with the
following functional form:
\begin{equation}
N(\phi_h) =N_0 \left( 1 + A^D_{\cos \phi}  \cos \phi_h +   A^D_{\cos 2\phi}
\cos 2\phi_h
 +  A^D_{\sin \phi} \sin \phi_h \right) 
\end{equation}
The contribution of the acceptance corrections to the systematic error was 
studied in detail.

\begin{figure}
\includegraphics[width=0.8\textwidth]{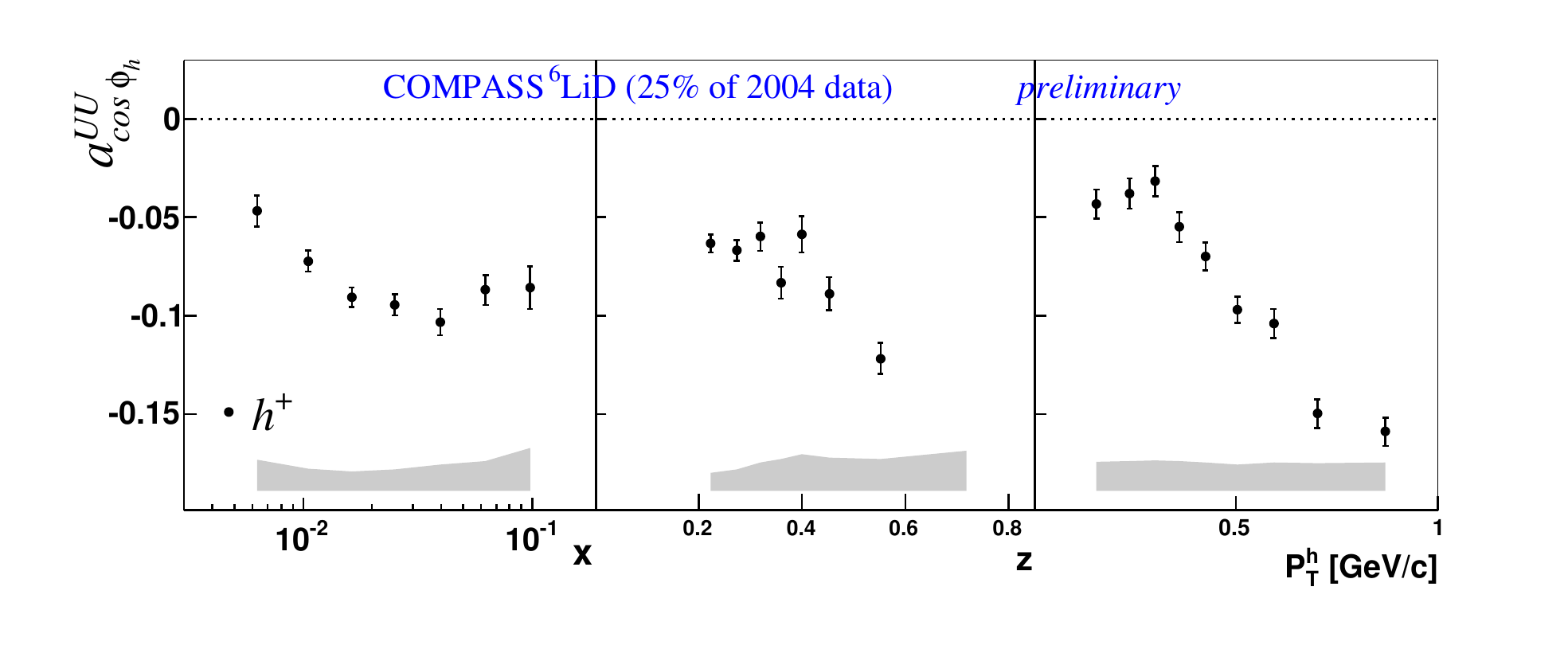}
\end{figure}
\begin{figure}
\includegraphics[width=0.8\textwidth]{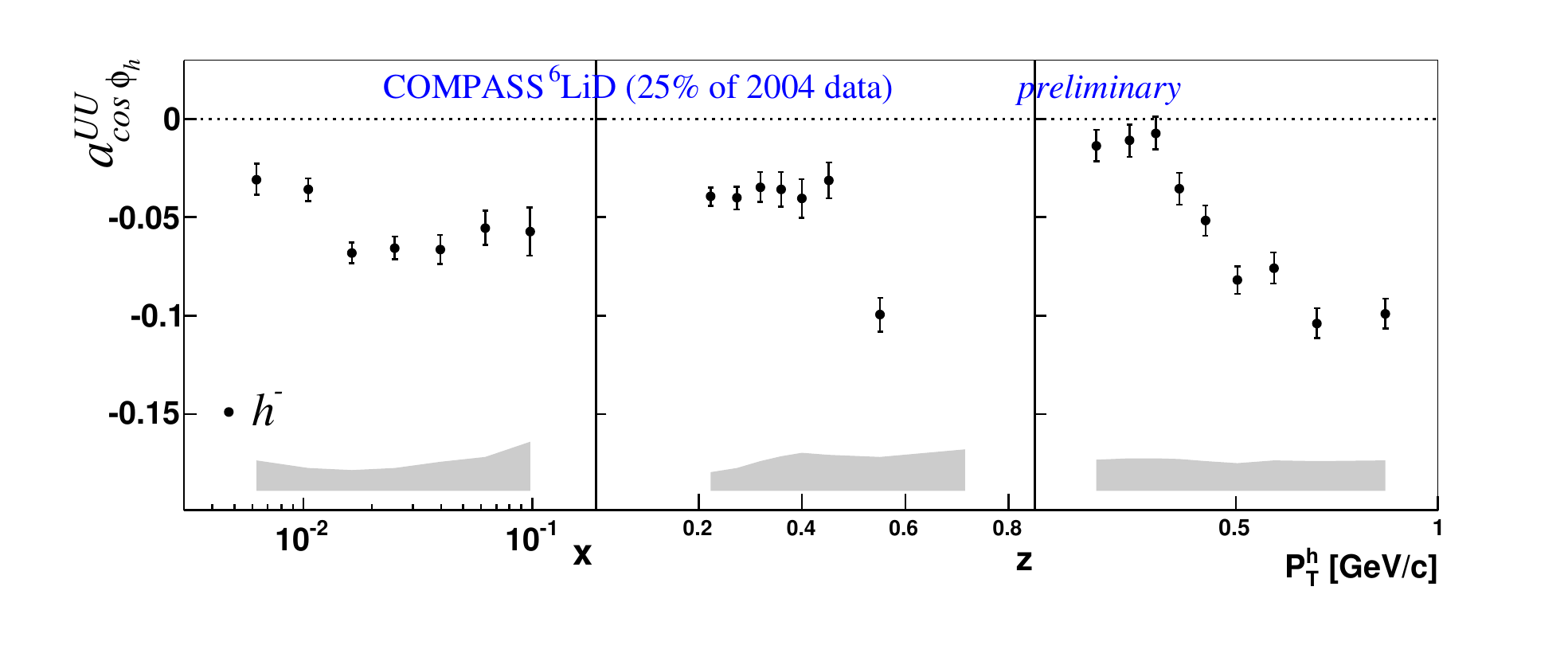}\vspace*{-1cm}
\caption{$\cos \phi_h$ asymmetries from COMPASS deuteron data
for positive (upper panel) and negative (lower
panel) hadrons. The bands indicate the size of the systematic uncertainty. 
}
\label{f:cosphi}
\end{figure}

The $\sin \phi_h$ asymmetries measured by COMPASS, not shown here,  are compatible
with zero, at the present level of statistical and systematic errors, over the
full range of $x$, $z$ and $p_T$ covered by the data.
The $\cos \phi_h$ asymmetries extracted from COMPASS deuteron data
are shown in Fig.~\ref{f:cosphi} for positive (upper panel) and negative (lower panel)
 hadrons, as a function of $x$, $z$ and $p_T$. The bands indicate the size
of the systematic error. The asymmetries show the same trend for positive and
negative hadrons with  slightly larger absolute values for  positive hadrons. 
Values as large as 30$-$40\% are reached in the last point of the $z$ range. Since the Cahn
effect does not depend on the hadron charge \cite{Cahn}, the difference between positive and
negative hadrons gives a hint to a non-zero
and flavor dependent Boer-Mulders distribution function. The measured asymmetries have been
compared  to model calculations, taking into account the Cahn effect for different values of the intrinsic
transverse momentum of the quarks $k_T$ \cite{Bacchetta2}.

\begin{figure}
\hspace*{-0.6cm}\includegraphics[width=0.6\textwidth]{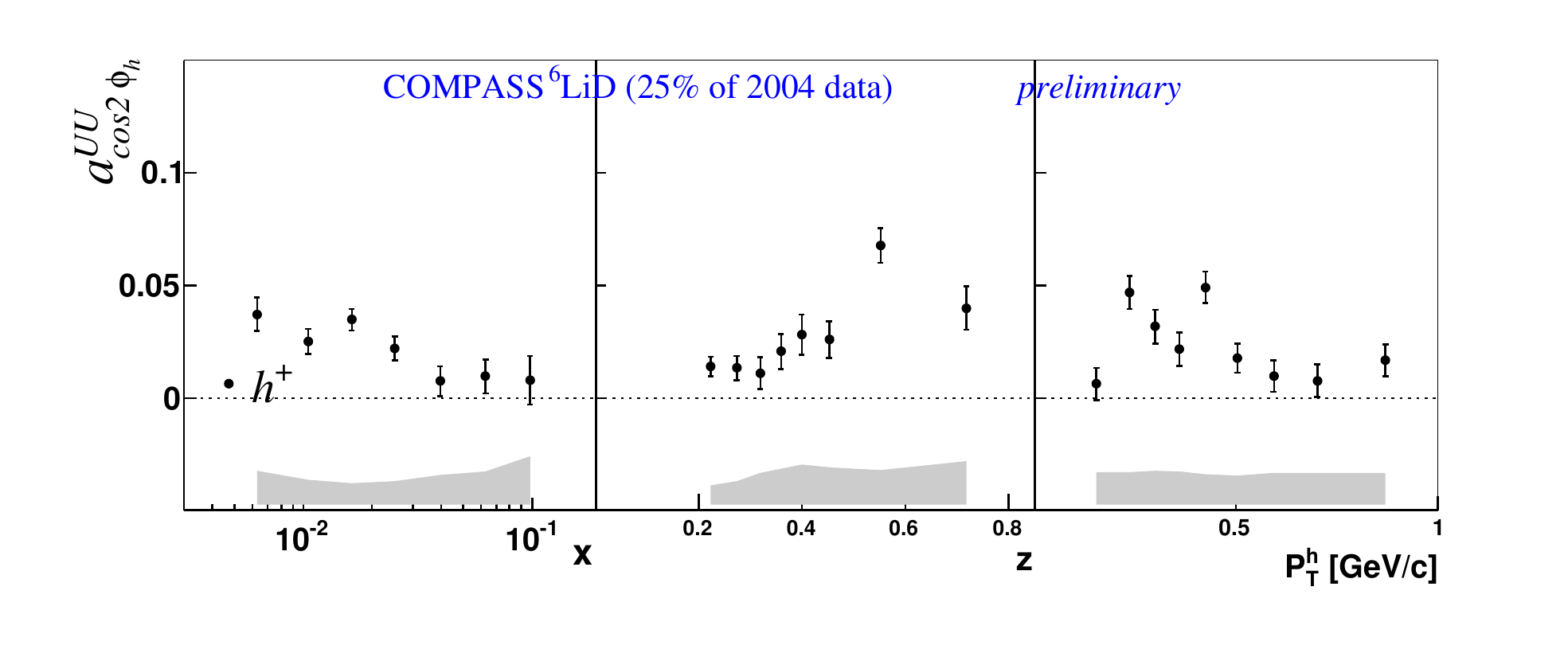}\hspace*{-0.6cm}
\includegraphics[width=0.6\textwidth]{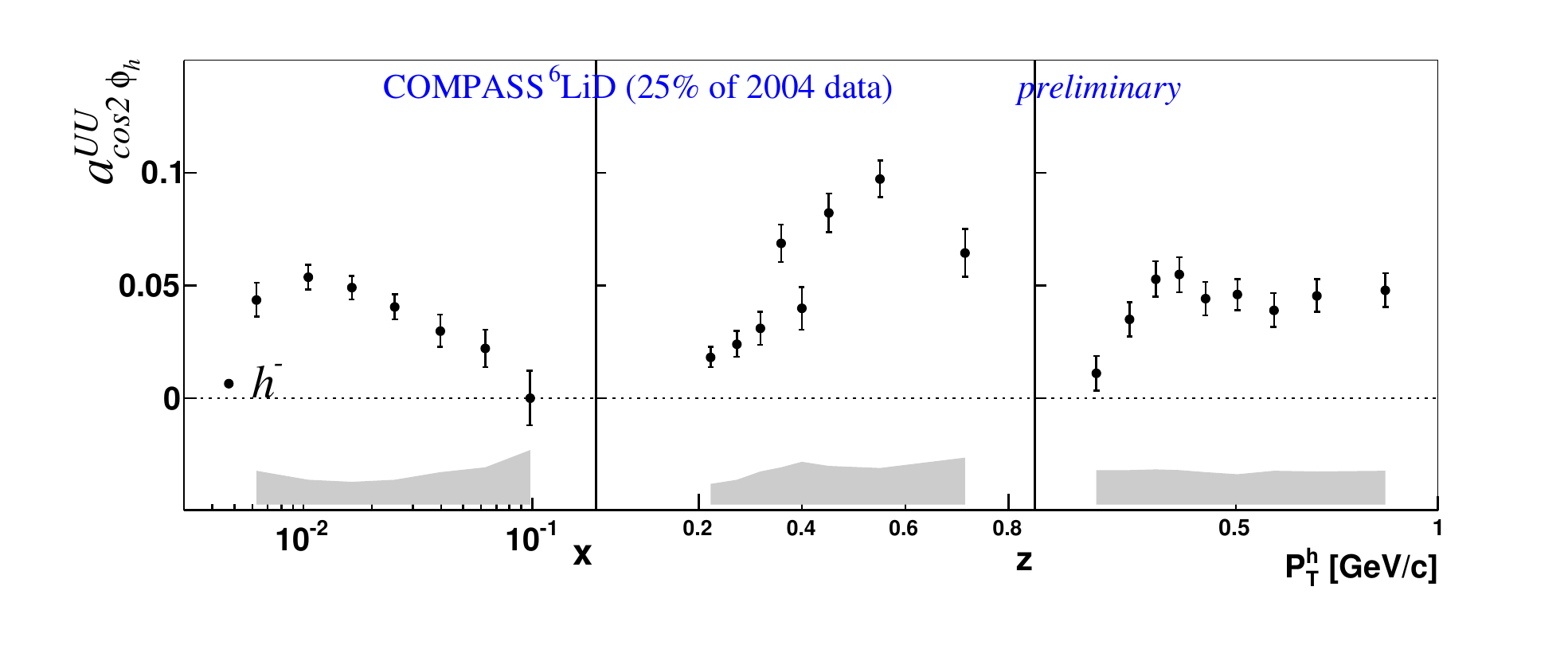}\vspace*{-1cm}
\caption{$\cos 2 \phi_h$ asymmetries from COMPASS deuteron data
for positive (left panel) and negative (right panel) hadrons. 
The bands indicate size of the systematic error.}
\label{f:cos2phi}
\end{figure}
The $\cos 2 \phi_h$ asymmetries are shown in Fig.~\ref{f:cos2phi}. The asymmetries are positive
for all hadrons and decrease with increasing $x$. The asymmetries show a similar trend for
positive and negative hadrons with larger absolute values for negative hadrons.

\section{Summary and Outlook}


The measured unpolarized azimuthal asymmetries on a deuteron target show large negative 
$\cos\phi_h$ moments for positive and for negative hadrons. The asymmetries are larger
for positive hadrons than for negative ones.  The $\cos 2\phi_h$ moments are all positive and
show a decreasing magnitude with increasing values of $x$.

\subsection*{Acknowledgments}

This work has been supported in part by the German BMBF.



\begin{thebibliography}{999}

\bibitem{Ralston} J.P. Ralston and D.E. Soper, Nucl. Phys. {\bf B 152}, 109
(1979).
\bibitem{Mulders} P.J. Mulders, R.D. Tangerman, Nucl. Phys. {\bf 461}, 197
(1997). 
\bibitem{Collins} J.C. Collins {\it et al.},  Nucl. Phys. {\bf B420}, 565 (1994).
\bibitem{Boer} D. Boer and P.J. Mulders, Phys. Rev. {\bf D57},  5780 (1998).
\bibitem{Cahn} R.N. Cahn, Phys. Lett. {\bf B78}, 269 (1978). 
\bibitem{COMPASS} V.Yu. Alexakhin {\it et~al.} [COMPASS collaboration] Phys.
Rev. Lett. {\bf 94}, 202002 (2005); E.S. Ageev {\it et~al.} 
[COMPASS collaboration] Nucl. Phys. {\bf B765}, 31 (2007); M.G. Alekseev {\it et al.} [COMPASS collaboration], Phys.
Lett. {\bf B692} (2010),  240. and M. Alekseev {\it et al.} [COMPASS
collaboration], Eur. Phys. J. {\bf C64} (2009), 171.
\bibitem{Experiment} P. Abbon {\it et~al.} [COMPASS collaboration], NIM {\bf
A577}, 455-518 (2007).
\bibitem{Bacchetta2}
A. Bacchetta {\it et al.}, JHEP {\bf 0702}, 93 (2007).
\bibitem{lepto}
G. Ingelman, A. Edin and J. Rathsman, Comp.Phys.Commun. {\bf 101}, 108 (1997).
\end{thebibliography}
\end{document}